\documentclass[12pt]{iopart}

\usepackage[colorlinks=true,linkcolor=blue,citecolor=blue,urlcolor=blue]{hyperref}
\usepackage{doi}
\usepackage{url}
\usepackage{graphicx}
\usepackage{amssymb}

\begin{document}

\title[]{LISA and the LISA Science Team}

\author{Anna Heffernan$^1$, Chiara Caprini$^{2,3}$, Neil Cornish$^4$, Jonathan Gair$^5$, Catia Grimani$^{6,7}$, Zoltan Haiman$^{8,9,10}$, Erin Kara$^{11,12}$, Nikos Karnesis$^{13,14}$, Valeriya Korol$^{15}$, Astrid Lamberts$^{16,17}$, , Nora L{\"u}tzgendorf$^{18}$, Guido M{\"u}ller$^{19,20,21}$, Gijs Nelemans$^{22,23,15}$, Antoine Petiteau$^{24}$, Elena Maria Rossi$^{25}$, Alberto Sesana$^{26,27,28}$, Joey Shapiro Key$^{29}$, Deirdre Shoemaker$^{30}$, Krista Lynne Smith$^{31}$, Stephen R. Taylor$^{32}$, Alberto Vecchio$^{33}$, 
William Joseph Weber$^{34}$ on behalf of the LISA Science Team}

\address{$^1$ Departament de F\'isica, Universitat de les Illes Balears, IAC3 – IEEC, Crta. Valldemossa km 7.5, E-07122 Palma, Spain}
\address{$^2$ Theoretical Physics Department, CERN, CH-1211 Gen\`eve, Switzerland}
\address{$^3$ D\'epartement de Physique Th\'eorique and Center for Astroparticle Physics,\\
Universit\'e de Gen\`eve, 24 quai Ernest  Ansermet, 1211 Gen\`eve 4, Switzerland}
\address{$^4$ EXtreme Gravity Institute, Department of Physics, Montana State University, Bozeman, Montana 59717 USA}
\address{$^5$ Max-Planck-Institut f\"ur Gravitationsphysik, Albert-Einstein-Institut, Am M\"uhlenberg 1, 14476 Potsdam-Golm, Germany}
\address{$^6$ Department of Pure and Applied Sciences, University of Urbino Carlo Bo, Via S. Chiara, 27, 61028 Urbino (PU), Italy}
\address{$^7$ INFN, Section in Florence, Via G. Sansone 1, 50019 Sesto Fiorentino (FI), Italy}
\address{$^8$ Institute of Science and Technology Austria (ISTA), Am Campus 1, Klosterneuburg 3400, Austria}
\address{$^9$ Department of Astronomy \& Astrophysics, Columbia University New York, NY 10027, USA}
\address{$^{10}$ Department of Physics, Columbia University New York, NY 10027, USA}
\address{$^{11}$ MIT Kavli Institute for Astrophysics and Space Research, 70 Vassar Street, Cambridge, MA 02139, USA}
\address{$^{12}$ Department of Physics, Massachusetts Institute of Technology, Cambridge, MA 02139, USA}
\address{$^{13}$ Department of  Physics, Aristotle University of Thessaloniki, Thessaloniki 54124, Greece}
\address{$^{14}$ Institute for Astronomy, Astrophysics, Space Applications and Remote Sensing, National Observatory of Athens, 15236 Penteli, Greece}
\address{$^{15}$ SRON Space Research Organisation Netherlands, Niels Bohrweg 4, 2333 CA Leiden, The Netherlands}
\address{$^{16}$ Laboratoire Lagrange, Observatoire de la Côte d’Azur, Université Côte d’Azur, CNRS, France}
\address{$^{17}$ Artemis, Observatoire de la Côte d’Azur, Université Côte d’Azur, CNRS, CS 34229, F-06304 Nice Cedex 4, France}
\address{$^{18}$ European Space Research and Technology Centre, Keplerlaan 1, 2200 AG Noordwijk, The Netherlands}
\address{$^{19}$ Max–Planck-Institut for Gravitation Physics (Albert-Einstein-Institut)}
\address{$^{20}$ Leibniz University Hannover, Hannover, Germany}
\address{$^{21}$ Department of Physics, University of Florida, Gainesville, USA}
\address{$^{22}$ Department of Astrophysics/IMAPP, Radboud University, P.O. Box 9010, NL-6500 GL Nijmegen, The Netherlands}
\address{$^{23}$ Institute of Astronomy, KU Leuven, Celestijnenlaan 200D, B-3001 Leuven, Belgium}
\address{$^{24}$ IRFU, CEA, Université Paris-Saclay, 91191, Gif-sur-Yvette, France}
\address{$^{25}$ Leiden Observatory, Leiden University, P.O. Box 9513, 2300 RA Leiden, The Netherlands}
\address{$^{26}$ Dipartimento di Fisica ``G. Occhialini", Universit{\'a} degli Studi di Milano-Bicocca, Piazza della Scienza 3, I-20126 Milano, Italy}
\address{$^{27}$ INAF - Osservatorio Astronomico di Brera, via Brera 20, I-20121 Milano, Italy}
\address{$^{28}$ INFN, Sezione di Milano-Bicocca, Piazza della Scienza 3, I-20126 Milano, Italy}
\address{$^{29}$ Physical Sciences Department, University of Washington Bothell,
18115 Campus Way NE, Bothell, WA 98011, USA}
\address{$^{30}$ Center for Gravitational Physics, Weinberg Institute, University of Texas at Austin, Austin TX, 78712, USA}
\address{$^{31}$ George P. and Cynthia Woods Mitchell Institute for Fundamental Physics and Astronomy, Texas A\&M University, College Station, TX 77843-4242, USA}
\address{$^{32}$ Department of Physics \& Astronomy, Vanderbilt University, 2301 Vanderbilt Place, Nashville, TN 37235, USA}
\address{$^{33}$ Institute for Gravitational Wave Astronomy \& School of Physics and Astronomy, University of Birmingham, Birmingham, United Kingdom}
\address{$^{34}$ Dipartimento di Fisica, Universit\`a di Trento and Trento Institute for 
Fundamental Physics and Application / INFN, I-38123 Povo, Trento, Italy}
\ead{anna.heffernan@uib.eu}

\begin{abstract}
LISA, the Laser Interferometer Space Antenna, due to launch mid-2035, is a large class space mission by the European Space Agency (ESA). In partnership with NASA and ESA-member states, ESA is on track to launch what is expected to be the first space-based gravitational wave detector. By hosting detectors in space, one gains access to a lower frequency band of gravitational wave sources and, with them, a plethora of new science. To maximise this scientific gain, ESA and NASA selected 20 scientists for the LISA Science Team to carry out and/or lead the necessary actions 
leading up to LISA's launch. We give a short overview and update of the LISA mission, its science objectives and related waveforms, as well as the work of the LISA Science Team as of April 2026.
\end{abstract}

%
%
\submitto{\CQG}
%
\maketitle
%
%


\section{Introduction}


\noindent At the time of adoption, January 2024, the LISA Mission Science Management Plan (SMP)~\cite{LISA:SMP} was established, describing how the different LISA stakeholders will work together for a successful mission. The plan included the formation of the LISA Science Team (LST), a group of twenty scientists selected by ESA and NASA and including a LISA Consortium representative, charged with maximising the scientific output of the LISA mission through various consultation and working group processes. The selected candidates were announced at the LISA Symposium in Dublin in July 2024 and started meeting regularly in September 2024. This mini review serves as a short summary of the LISA mission and its science objectives, suitable for newcomers to the LISA mission, as well as a summary of the work carried out by the LST to-date, as they pass 18 months of regular meetings. Sec.~\ref{sec:mission} gives an overview of the mission and stakeholders, Sec.~\ref{sec:science} summarises the science objectives of LISA while Sec.~\ref{sec:LST} describes the work of the LST.


\section{The LISA Mission} \label{sec:mission}

\begin{figure}
    \centering
    \includegraphics[width=\textwidth]{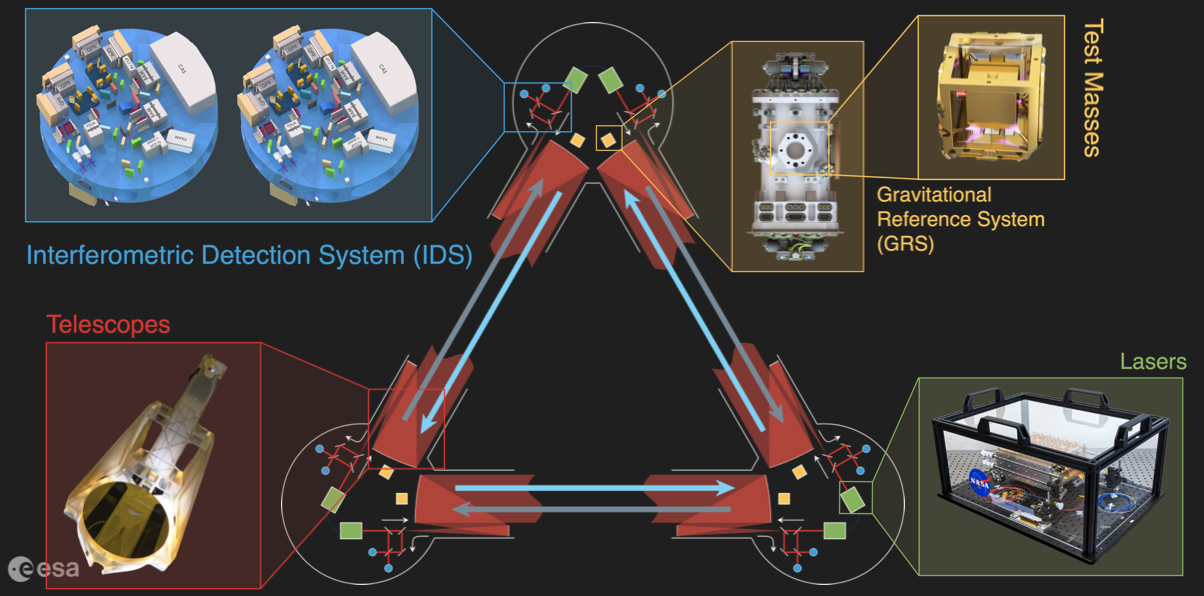}
    \caption{Schematic of the LISA detector comprising of three spacecraft with six null connections. Each spacecraft will have two telescopes, two lasers, two interferometric detection systems (IDSs) and two gravitational reference systems (GRSs)}
    \label{fig:detector}
\end{figure}

LISA will detect gravitational waves (GWs) in the mHz range, complementing 
current and future ground detectors; the mission and its objectives are 
described in the 
LISA Redbook~\cite{LISA:2024hlh}. 
A 
constellation of 3 spacecraft (SC) forming an 
almost equilateral triangle, inclined by 60 degrees to the orbital plane, 
LISA's 
centre of mass trails the Earth's orbit  (via 3 slightly 
inclined orbits around the sun). Lasers, emitting from each SC to the other two, form six null connections and allow tracking of the inter-SC distances. Each connection (two per SC) 
entails a test mass (TM) within a gravitational reference system (GRS), an interferometric detection system (IDS) that includes an optical bench (OB), a telescope and a laser (Fig.~\ref{fig:detector}). The proper distance between TMs (shielded within each SC) is then measured by combining TM to OB and OB to OB interferometry data.

The TM, a 46~mm, roughly 2~kg cube of gold-platinum, 
and 
surrounding hardware 
form 
the GRS.  The TM is enclosed by a conductive electrode housing that serves as an electromagnetic shield and defines the free-falling environment for the TM.  It includes electrodes that simultaneously allow a 6 degree-of-freedom position readout and application of nanoNewton-level electrostatic forces, which together allow the dynamical control of the relative TM-SC motion and alignment of the TM to the optical system. It has a launch-lock device to protect the TM during launch and separate device for positioning and releasing the TM in orbit.   
A UV illumination system can neutralize the charge accumulated on the TMs  by galactic cosmic rays and solar energetic particles~\cite{Grimani:2022bkr, Wass:2022igo, 2025A&A...700A.102D},
while the vacuum chamber around the TM is maintained via a venting duct to space; the SC also houses gravitational balance masses required to compensate the self-gravity of the SC and instrumentation.  The GRS system has been fully tested in the LISA Pathfinder mission, demonstrating fm/s$^2$/Hz$^{1/2}$ free-fall and the LISA TM acceleration requirements across the entire frequency band, in addition to various key measurement science aspects of LISA \cite{lpf_noise_2016,lpf_noise_final}. 

The telescopes both transmit and receive collimated beams between the neighbouring SC and their OBs; separation of the beams is ensured via different polarisations. Each OB consists of 3 interferometers: the inter-satellite interferometer tracks the relative motion (includes angular) between the OBs on each SC, the TM interferometer measures motion between the OB and the TM (again including angular), and the reference interferometer tracks relative phase fluctuations between the two lasers on board (one in each IDS); this allows phase-locking the lasers used in adjacent arms of the constellation. 
Each laser outputs a power of 2W, yet only a few 100 picowatts are received due to the divergence of the beams travelling 
between SC. The weak received beam is interfered with a small fraction of the outgoing laser beam, with the frequency difference or "beat note" measured by a dedicated phasemeter.  This main science signal, limited by shot noise in the weak incoming beam, contains the micro-Hz frequency modulation from the gravitational wave tidal acceleration on the LISA constellation. 

The telescope, laser, and UV light source for TM photoelectric discharge (the charge management device or CMD) are being developed by NASA with all other components being produced by ESA and its member states. We are currently in phase B2; the prime contractor has been selected (OHB), payload preliminary design review is in progress, and payload critical design review will begin shortly. 
The mission at this phase (implementation) 
has four structures
: the instrument and project team, that is the ESA LISA Project, Performance and Operations Teams with all instrument providers; the science ground segment (SGS) (European Distributed Data Processing Centre, NASA SGS and Science Operations Centre); the LISA Science Team (LST) and its working groups; and the scientific community (including the LISA Consortium) that interacts with both the SGS and the LST. Details of the mission setup are in the Science Management Plan (SMP)~\cite{LISA:SMP}. 

\section{LISA Science} \label{sec:science}


\begin{figure}
    \centering
    \includegraphics[width=\textwidth]{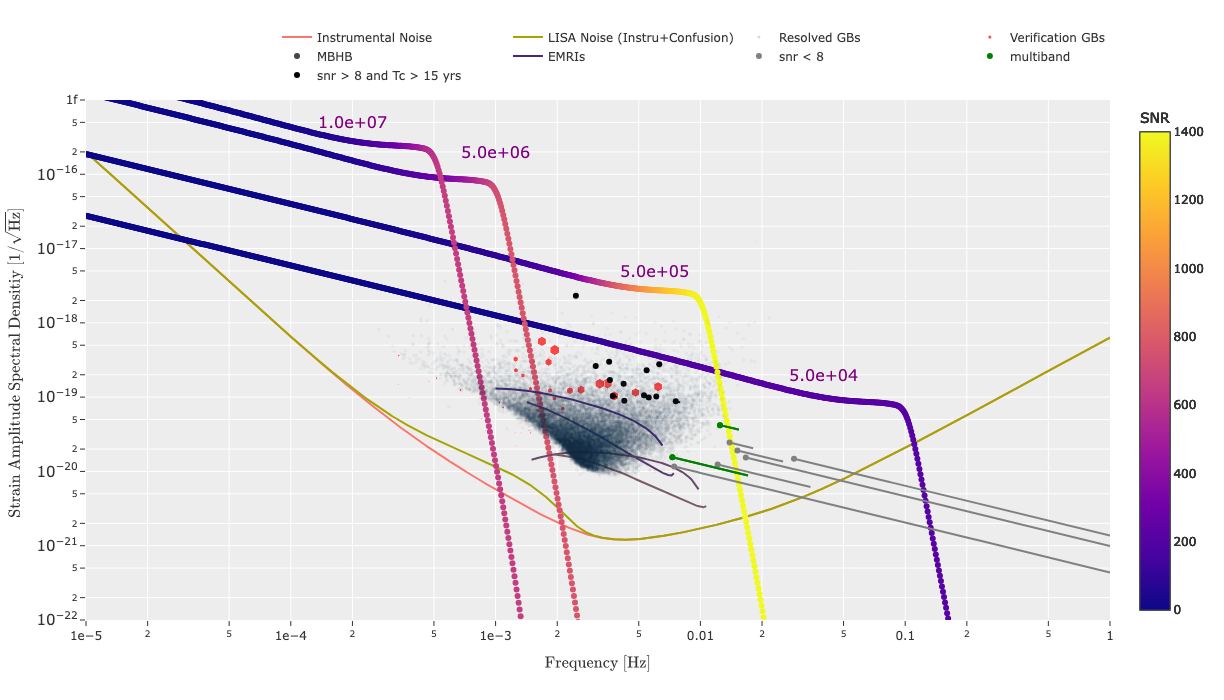}
    \caption{LISA sensitivity plot as produced by the interactive figures of merit website \cite{FoM:2024active}, developed by the LISA Consortium for the LISA Redbook~\cite{LISA:2024hlh}. LISA's instrumental noise combined with the expected astrophysical confusion from the galactic-binary foreground dictate LISA's sensitivity. Massive black hole binaries (MBHBs), resolvable galactic binaries (including verification binaries), multiband stellar-mass black hole binaries (sBHBs) and the four dominant modes of a single extreme mass-ratio inpsiral (EMRI) are illustrated for a mission duration of 4.5 years.}
    \label{fig:sensitivity}
\end{figure}
The LISA science objectives (SOs), which we summarise here, are 
described in detail
in the LISA Redbook~\cite{LISA:2024hlh}. 
In-depth reviews of the 
Waveform modelling, Astrophysics, Cosmology and Fundamental Physics can be found in a series of White Papers produced by the LISA Consortium~\cite{LISAConsortiumWaveformWorkingGroup:2023arg, LISA:2022yao, LISACosmologyWorkingGroup:2022jok, LISA:2022kgy}.


\subsection{SO1: Study the formation and evolution of compact binary stars and the structure of the Milky Way Galaxy} 


\begin{figure}
    \centering
    \includegraphics[width=0.6\textwidth]{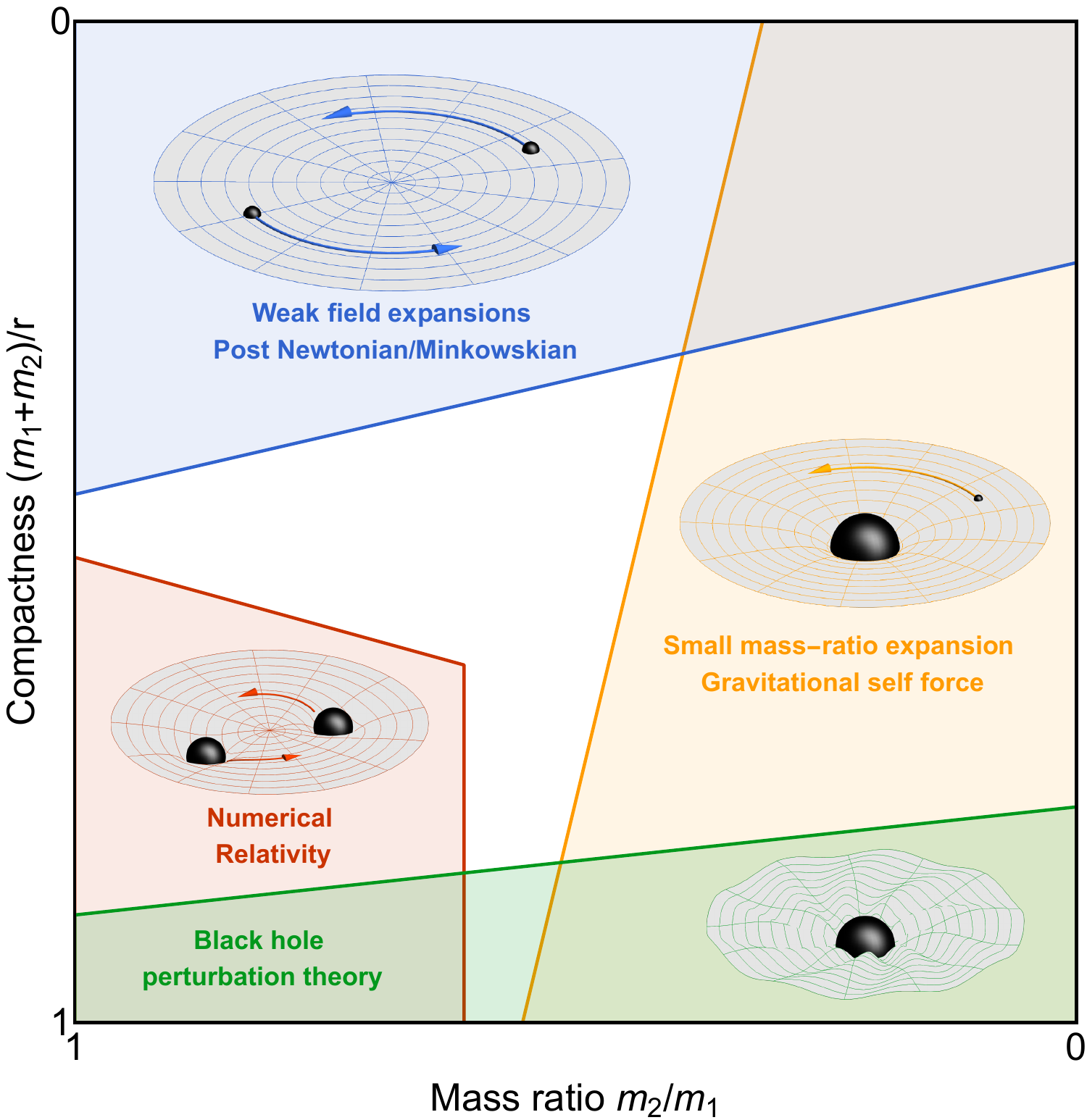}
    \caption{Parameter space coverage for different waveform source-modelling techniques taken from~\cite{LISAConsortiumWaveformWorkingGroup:2023arg}. Post-Newtonian (PN) approximations perturb using the slowness of the system or velocity compared to the speed of light, and hence breaks down as the binary tightens. Numerical relativity (NR), where one numerical solves the partial differential equations of General Relativity, becomes both costly and inefficient with more extreme mass ratios or long inspirals. Gravitational Self-Force (GSF) perturbs the Einstein field equations in the mass-ratio and hence breaks down for more comparable masses. Black hole perturbation theory involves a perturbations of an isolated black hole to capture the post-merger ringdown. Effective one body (EOB) and Phenomenological (Phenom) models combines these different models to provide smooth waveform models covering inspiral through merger and ringdown across larger parts of the parameter space, while NRsurrogate interpolates over the NR densely populated part of the parameter space.}
    \label{fig:parameterSpace}
\end{figure}
This mostly refers to double white dwarfs (DWDs), of which about $10^4$ will be individually detectable by LISA, seen as the navy cloud of dots in Fig.~\ref{fig:sensitivity}; in turn, several hundred of these are expected to have electromagnetic (EM) counterparts enabling multi-messenger studies~\cite{Breivik:2017jip}. Tens of EM signals have already been identified as DWDs detectable by LISA, the so-called verification binaries~\cite{Kupfer:2023nqx}, which will assist in the scientific verification of the LISA data (red dots in Fig.~\ref{fig:sensitivity}). Many more $(\sim10^7)$ DWDs are expected to also emit in the LISA range building a stochastic Galactic foreground~\cite{Boileau:2021sni}. Binaries consisting of either or both neutron stars (NSs) and black holes (BHs) should also be detectable from within our Galaxy, however, they will likely be hard to identify in the LISA data and will occur at much lower numbers (tens to hundreds).

In modelling, one only 
considers the inspiral as 
these binaries are at large separations and will not merge in the LISA band; thus, the 
post-Newtonian (PN) approximation~\cite{Blanchet:2013haa} 
is used (Fig.~\ref{fig:parameterSpace}). In fact, due to weak GW emission at 
this stage, most signals are expected to be quasi-monochromatic, with limited frequency evolution during the mission. 
A mass-transfer phase can occur in LISA-band DWDs
, which determines the 
binary's fate: either a merger or stable mass transfer that counteracts GW radiation and widens the binary.
Tidal effects are also important
; their interplay with both mass transfer 
and GW emission is poorly understood and can affect 
the binary's final state~\cite{Marsh:2003rd}. For systems with NSs/BHs, close binaries inform us of their initial kick velocity 
from the individual NS/BH formation 
during supernova 
(already seen by EM observations for NSs); high kick velocities will tend to disrupt binaries and this will be observed in their distribution~\cite{Giacobbo:2018etu}. Indeed, the sheer number of expected detections will not only allow us to make statistical reasoning on all the above but also enables us to map out the Milky Way mass distribution and inform merger rates. 


\subsection{SO2: Trace the origins, growth and merger histories of massive Black Holes across cosmic epochs}


\begin{figure}
    \centering
    \includegraphics[width=0.9\textwidth]{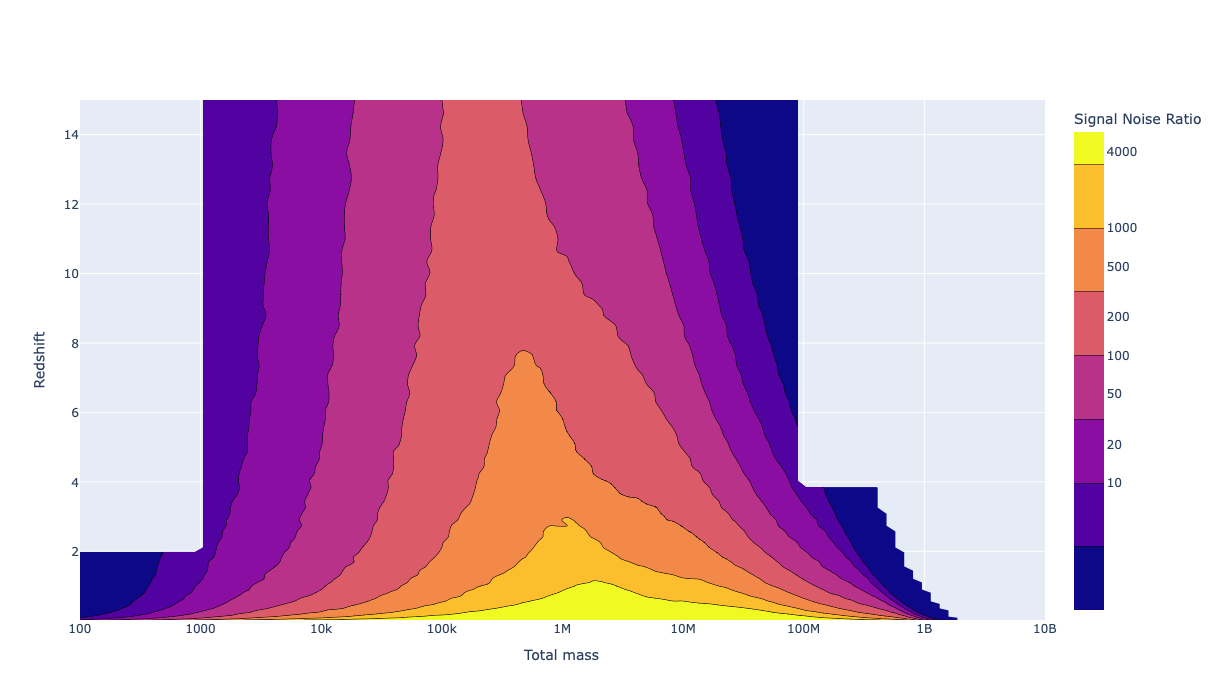}
    \caption{A waterfall plot for massive black hole binaries (MBHB) for a 4.5 years mission, as produced by the interactive figures of merit website \cite{FoM:2024active}, developed by the LISA Consortium for the LISA Redbook~\cite{LISA:2024hlh}. The x-axis shows multiples of solar mass with $M=10^6$ and $B=10^9$ equating to a million and billion respectively. This illustrates the extremely high SNR signals expected for MBHBs at large redshifts, $z$.}
    \label{fig:waterfall}
\end{figure}
Massive BH binaries (MBHBs: masses ~$\sim 10^6 - 10^9 M_\odot$)
, are well outside the scope of ground detectors. LISA will see MBHBs out to an arbitrarily large redshift ($z \sim 12$ and beyond, see Fig.~\ref{fig:waterfall}) as well as less massive binaries involving intermediate mass BHs (IMBHs: masses $\sim10^2 - 10^5 M_\odot$). Little is known about IMBHs; 
only a few on the extremes of their mass range have ever been detected~\cite{LIGOScientific:2025rsn, Lin:2018dev, 2020ARA&A..58..257G}, allowing little insight into their origin and evolution spanning these masses. LISA's ability to detect them will create a new pool of knowledge. Meanwhile massive BHs (MBHs) have long been confirmed in both the present and early universe by EM observations, including accreting~$10^6 M_\odot$ MBHs at redshifts~$4<z<10$~\cite{Maiolino:2023bpi, Banados:2017unc}. 
LISA's sensitivity to~$10^3 - 10^7 M_\odot$ binaries at such redshifts will inform theories on BH growth and their host galaxies across the cosmos, in particular IMBHs at the epoch of MBH formation ~$z>10$, which is outside all current telescope's abilities
. This in turn will not only 
unveil insights into
their origin, 
and growth, but also inform spin distributions and merging rates. As the signals can stay in the detector from days to weeks (mass dependent), one can alert the global network of EM telescopes to search for multi-messenger signals. This could lead to environmental information on accretion as well as a wealth of complementary data (the only GW-EM multi-messenger to-date, GW170817, led to a groundbreaking number of scientific observations~\cite{LIGOScientific:2017ync}).

In detecting 
MBHBs, a problem arises: LISA
's sensitivity allows jarring SNRs (1000s), exasperating the infamous 
global fit problem. LISA will see all sources from all directions simultaneously (modulated by LISA's orbital motion); and, therefore, one 
must
systematically identify and remove signals from the data. SNRs~$\sim1000$ require waveforms of 
unparalleled
accuracy for removal without remnants poisoning lower-SNR signals. Comparable-mass binary 
models 
used by current ground detectors~\cite{LIGOScientific:2025yae} combine PN and numerical relativity (NR) (Fig.~\ref{fig:parameterSpace}), 
balancing speed, accuracy and parameter space coverage, e.g. effective one body~\cite{Ramos-Buades:2023ehm, Akcay:2018yyh}, phenomenological~\cite{Colleoni:2024knd, Thompson:2023ase} and NR surrogates~\cite{Varma:2019csw}. Neither their current accuracy nor parameter space coverage 
is good enough for LISA MBHBs~\cite{LISAConsortiumWaveformWorkingGroup:2023arg}.


\subsection{SO3: Probe the properties and immediate environments of Black Holes in the local Universe using extreme mass-ratio inspirals and intermediate mass-ratio inspirals}


Compact objects are predicted to orbit and merge with MBHs and IMBHs
. Stellar-mass BHs~($5-10^2 M_\odot$) inspiralling into a MBH, known as extreme mass-ratio inspirals (EMRIs), are expected to occur in galaxy centres. A stellar-mass BH merging with an IMBH or an IMBH with a MBH, both called intermediate mass-ratio inspirals (IMRIs), are identified as light or heavy IMRIs respectively. Light IMRIs are expected to arise in dense star clusters and dwarf galaxies. Both EMRIs and IMRIs generate many GW cycles ($\sim 10^5$) in the LISA band, 
enabling tight constraints on the primary's spin~($\sim 10^{-5}$) 
as well as the secondary's eccentricity and inclination, while masses will be measured to~$\sim 10^{-2}$. These constitute detections of MBHs and IMBHs accordingly and 
so 
will provide new information 
on the population, parameters and growth mechanisms. 
Spin, eccentricity and inclination will inform formation channels~\cite{Hopman:2005vr, ColemanMiller:2005rm, Volonteri:2004cf}, while the environment (accretion disk~\cite{Duque:2025yfm, HegadeKR:2025rpr}, multiple bodies~\cite{Pan:2023wau}) may also leave an 
imprint on the waveforms.

In modelling EMRIs and IMRIs
, PN struggles as the binary tightens
, while NR slows for differing body sizes. The self-force (SF) program (Fig.~\ref{fig:parameterSpace}), which perturbs in the mass-ratio, has emerged as the primary modelling approach. The first post-adiabatic SF inspiral-only waveform
~\cite{Wardell:2021fyy} showed 
consistency
with NR for mass-ratios as low as 10, confirming its application to IMRIs as well as EMRIs. Current models cover only circular inspirals, with a spinning secondary~\cite{Mathews:2021rod} and small primary spin ($\chi<0.1$)~\cite{Mathews:2025txc}, with promising merger-ringdown 
developments
~\cite{Kuchler:2025hwx, Honet:2025dho}. Generic (spinning, eccentric, inclined) post-adiabatic SF waveforms are required for parameter estimation of EMRIs and IMRIs~\cite{Burke:2023lno}; fast (less accurate) adiabatic waveforms for spinning eccentric systems have been developed in the meantime~\cite{Chapman-Bird:2025xtd} for use in astrophysical studies and for data analysis developments. 
Combining NR with SF~\cite{Wittek:2024pis} or 
PN with SF fluxes~\cite{Honet:2025lmk} may also 
yield generic waveforms.


\subsection{SO4: Understand the astrophysics of stellar-mass Black Holes}


Over 200 stellar-mass BHBs (sBHBs) have been detected by 
the LIGO-Virgo-KAGRA Collaboration (LVK) of ground detectors~\cite{LIGOScientific:2025pvj}, which is expected to reach~$\sim10^4$ by the time LISA flies
. Ground detectors catch binaries merging, when most have circularised; LISA will see sBHBs earlier in their inspiral (green and gray lines in Fig.~\ref{fig:sensitivity}) with sensitivity to eccentricity (informs formation channels) and environmental effects. BHs that grew together, from massive stars of a stellar binary collapsing, expect low eccentricity and aligned spins. Binaries formed via dynamical capture 
will generate a more random distribution of spins in eccentric orbits. LISA may also detect higher-mass sBHBs, like GW190521~\cite{LIGOScientific:2020iuh}, that sit in the theoretical BH pair-instability mass-gap (a range of masses for which BHs can not form directly from star collapse~\cite{Woosley:2021xba}). BHs in this range form via 
hierarchal mergers~\cite{Gerosa:2021mno} or a combination of mergers and accretion within a disk~\cite{Tagawa:2020qll, Bartos:2025pkv}, with each having their own signatures in the waveforms~\cite{Inayoshi:2017hgw}, and hence environmental information. In addition, LISA may see a sBHB that is later detected by ground detectors, allowing a multi-band detection and the ability to send early alerts (months prior to merger), with sky position and expected merger time (down to seconds), to both the EM community (to search for EM counterparts) and the ground detectors (ensure they are taking data). Multiband detections are notable probes in testing Einstein's relativity~\cite{Baker:2022eiz}
, the subject of 
SO5. In modelling sBHBs, due to their comparable mass, one may use the same waveform families and techniques as MBHBs (where generic waveforms are also required).


\subsection{SO5: Explore the fundamental nature of gravity and Black Holes}


Beyond General Relativity theories not only allow for additional fields (e.g. scalar hair) but also polarisations, to which LISA is sensitive \cite{Tinto:2010hz}. General relativity dictates that the postmerger ringdown modes of a BHB are completely described by the Kerr BH's mass and spin (no-hair theorem); one can therefore infer these properties if one mode is measured, while further modes enable one to test or further constrain this prediction \cite{Berti:2005ys}. The large SNRs expected in the LISA data of MBHBs makes them the perfect lab for such tests; indeed with 3 or more modes of SNR of 8 or higher, one can confirm the Kerr nature of the remnant within $1 - 10$\%~\cite{Toubiana:2023cwr} 
(and rule out more exotic objects e.g. boson stars, fuzzballs). One can also leverage the long inspirals of EMRIs and multiband sBHBs to probe fundamental physics.  The $10^4-10^5$ orbits of EMRIs enable tight constraints on the mass (detector frame), spin and quadrupole moment of the primary Kerr BH~\cite{Babak:2017tow}. Deviation from a Kerr BH by either body could be detectable or if not, constrained. Multiband sBHBs cover a large frequency range, also enabling high precision in detecting or contraining deviations from General Relativity~\cite{Barausse:2016eii, Carson:2019rda}. Lastly, deviations in GW propagation properties can have implications for fundamental physics, be it speed (dark energy models of frequency dependent GW speed~\cite{deRham:2018red}), phase (provides constraints on graviton mass~\cite{Perkins:2020tra}), or amplitude (dark energy dampening over large distances~\cite{LISACosmologyWorkingGroup:2019mwx}).


\subsection{SO6: Probe the rate of expansion of the Universe with standard sirens}


GW signals enable direct measurement of the luminousity distance to their source, whereas EM radiation allows redshift measurements. A multi-messenger signal of both EM and GW emissions from the same source can thus provide a ``standard siren'', measuring both to probe the expansion rate of the Universe and infer cosmological parameters~\cite{Schutz:1986gp}. ``Bright sirens'' are the ideal scenario, where one receives a direct EM emission from the binary, similar to the binary neutron star event, GW170817~\cite{LIGOScientific:2017ync}. For MBHBs, this involves identifying the host galaxy, which is possible either when the sky-localisation from the GW emission is very precise or when the binary position can be inferred through its EM emission e.g. a radio flare~\cite{Mangiagli:2022niy}. 
LISA's senstivity to MBHBs (Fig.~\ref{fig:waterfall}) overlaps with future EM observatory capabilities like Athena, SKA, ELT and Vera Rubin Obervatory, in the interval $1.5 \lesssim z \lesssim 6 $, allowing one to constrain the Hubble rate at high redshift, e.g. $H (z\sim2)$ within a 10\% accuracy or less~\cite{Mangiagli:2022niy, Mangiagli:2023ize, Speri:2020hwc}. EMRIs at more limited redshifts, on the other hand, can act as ``dark sirens''; sources with precise enough sky localisation, in both angular position and luminousity distance, that one can infer their redshift, statistically match them to galaxy catalogues and thus constrain the Hubble constant to a few percent at $z<3$ (\cite{Laghi:2021pqk} assumes fiducial EMRI population and complete galaxy catalogs up to $z=1$).


\subsection{SO7: Understand stochastic GW backgrounds and their implications for the early Universe and TeV-scale particle physics}


\begin{figure}
    \centering
    \includegraphics[width=0.9\textwidth]{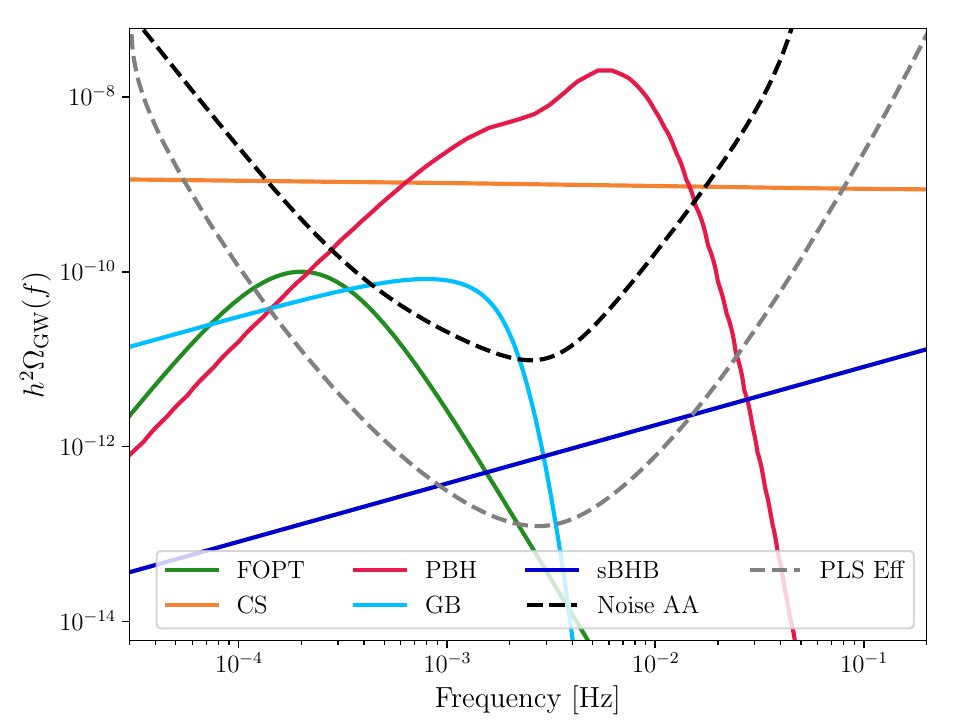}
    \caption{The LISA sensitivity curve with the various stochastic GW background (SGWB) signals illustrated, taken from ~\cite{LISA:2024hlh}. This can be compared to Fig.~\ref{fig:sensitivity}, where the LISA instrumental noise (dashed black line) and astrophysical galactic binary foreground (light blue from~\cite{Karnesis:2021tsh}) are combined to give the overall sensitivity of LISA. Here we see astrophysical SGWBs from sBHBs (dark blue from~\cite{Babak:2023lro} assuming GWTC-3 population constraints~\cite{KAGRA:2021duu}) and cosmological SGWBs from  first order phase transitions (green), cosmic strings (orange, assuming a string tension that would account for the recent SGWB detection by pulsar timing arrays~\cite{EPTA:2023xxk}) and primordial BHs (red, for the totality of dark matter~\cite{Bartolo:2018rku}). It should be noted that in illustrating sensitivity to stochastic backgrounds, it is more informative to compare signals to the effective power-law sensitivity~\cite{Thrane:2013oya} integrated for mission duration (4 years in dashed grey). Recent studies have revealed an additional astrophysical SGWB from extra-galactic WDs missing from this figure~\cite{Boileau:2025jkv}}
    \label{fig:SGWB}
\end{figure}
A stochastic background of GWs, like a roar from a crowd, is constituted by many similar overlapping signals that are individually unresolvable, but combined are detectable. In the LISA band, backgrounds of both astrophysical and cosmological origin are predicted. In LISA's sensitivity curve, Fig.~\ref{fig:sensitivity}, one can see the impact of an astrophysical foreground predicted from galactic binaries~\cite{Boileau:2021sni}, while Fig.~\ref{fig:SGWB} illustrates other detectable stochastic backgrounds. It is also possible that LISA will see astrophysical backgrounds from other close sources, e.g. extra-galactic WDs~\cite{Boileau:2025jkv}, sBHBs~\cite{Babak:2023lro} and EMRIs~\cite{Bonetti:2020jku}. 
Although these are not visible in the current LISA sensitivity curve, Fig.~\ref{fig:sensitivity}, this does not mean LISA and the relevant data pipelines will not search or accommodate them. Detection of any astrophysical background will inform the relevant population models and formation channels.

A cosmological stochastic GW background (SGWB), similar to the Cosmic Microwave Background, consists of radiation from the earliest moments of our universe. There are several well-motivated scenarios that could lead to a detectable cosmological background in the LISA band. For example, if the spontaneous breaking of the electroweak gauge symmetry (which separated the electroweak force into the distinct electromagnetic and weak nuclear forces) is not a smooth crossover, but instead proceeds as a first-order phase transition (more abrupt) via bubble nucleation, the resulting SGWB could be detectable by LISA~\cite{Caprini:2019egz}. Cosmic strings could also result from phase transitions at the high energies of the early universe and lead to a detectable SGWB in LISA~\cite{Auclair:2019wcv}, as could 
a primordial BH population that accounts for the totality of dark matter~\cite{Bartolo:2018rku}. A lack of detection will constrain these models far beyond the current state of the art, while any detection would be a groundbreaking discovery, offering us the rarest of glimpses into the universe's earliest moments.


\subsection{SO8: Search for GW bursts and unforeseen sources}


In addition to forming a cosmological SGWB, cosmic strings can also be the source of burst sources. Intersections of strings as well as points on the strings momentarily moving at the speed of light, known as kinks and cusps respectively, can generate a burst signal. The expected emission is concentrated in a cone, of which the opening angle is inversely proportional to the frequency ($f^{-1/3}$ to be precise)~\cite{Damour:2000wa, Damour:2001bk}, thus, these signals are more likely to be seen by LISA than its groundbased counterparts. Similar to the scenario of SGWB, a lack of detection would constrain the string tension. Along with bursts from cosmic strings, as is the case for any new window onto our universe, there is the possibility of unknown or unanticipated burst sources in LISA. The prospect of detecting an unknown source is being anticipated whilst developing the LISA data pipelines and would constitute an extraordinary finding.


\section{The LISA Science Team (LST)} \label{sec:LST}


The LST, 20 scientists with a wide range of expertise from institutes across Europe and the USA, were selected over three calls
. As per the SMP~\cite{LISA:SMP}, the initial 18, announced in July 2024, included 6 from a NASA call, 11 from an ESA call and the LISA Consortium nominated 
representative, in areas of Astrophysics (Neil Cornish, Erin Kara, Valeriya Korol, Astrid Lamberts, Gijs Nelemans, Elena Maria Rossi, Alberto Sesana, Joey Shapiro Key, Krista Lynne Smith, Alberto Vecchio), Cosmology (Chiara Caprini), Data Analysis (Nikolaos Karnesis, Antoine Petiteau, Stephen Taylor), Instrumentation (Guido M{\"u}ller, William Joseph Weber) and Waveforms (Anna Heffernan, Deirdre Shoemaker). These were later joined in April 2025 by 2 ESA-selected complementary scientists in Space Weather (Catia Grimani) and Multi-Messenger Astrophysics (Zoltan Haiman), while in October 2025, transfer of the Consortium representative role from Gijs Nelemans to Jonathan Gair began. The LST goals~\cite{LISA:SMP} are towards maximising the science return of LISA, including communications and access. Occasionally, an urgent matter leads to a task force formation; this was the case for LISA input into the European Strategy for Particle Physics, where a team (including external members) delivered a report within weeks due to a tight deadline~\cite{Caprini:2025mfr}. More defined longterm goals are tackled via working groups (WGs), which can invite external members. There are currently 6 WGs: Alerts, Authorship, Communications, Figures of Merit (FoM), L3 Catalogue and Science Topical Panels (STPs).


\subsection{The Alerts Working Group}


Initialized at the LST face-to-face in December 2025, this WG  is 
chaired by ZH and VK. The goals are to provide inputs and specifications to the SGS in developing a pipeline for issuing alerts; design recommendations for SOC on when and how to operate this pipeline and issue alerts; and connect with communities outside LISA to ensure awareness and lead-time needed for triggered EM observations. 


\subsection{The Authorship Working Group}


This WG, chaired by NC and JG, aims to create a set of criteria as well as a procedure to populate the heritage and member author lists as described in the SMP \cite{LISA:SMP}. The heritage author lists those who have made a significant contribution to the mission and has no expiration date. The member author list comprises people working on the mission at the time of science operations and has a roll-off period of 2 years. To-date, a survey has been done among the LST members and information has been collected on criteria used by other missions / instruments for authorship. Descriptions of the member and heritage author lists have been formulated and discussed with the full LST.  Based on this feedback, the description of the heritage list has been organized under the two categories, {\em Founders} and {\em Builders}.  Next steps will be to create drafts of the criteria for membership and the selection procedure.


\subsection{The Communications Working Group} 


The Communivations WG, chaired by AH and KLS, aims to ensure smooth communications, both internal and external
, formally and informally (via mutual members). Internal 
connects the different WGs, LST members, and project scientists. External refers to interested entities, including the Distributed Data Processing Centre (DDPC members: JG), NASA Science Ground Segment (NSGS: external member Ann Hornschemeier Cardiff of NASA), Lisa Consortium (AH, JG, VK, JSK), Gravitational Wave International Committee (GWIC: JG, AH, JSK) and other such stakeholders.
Several lines of external communication have been established: the LST has an official email (LISAScienceTeam@esa.int), moderated by AH and KLS; 
an FAQ list 
has been created, which will go live on the ESA LISA website shortly (with other prepared material by this WG in collaboration with the ESA project team); close ties to the Consortium communications is set up via shared members VK and JSK; procedure guidelines have been created for recruiting temporary task forces or WG members as well as receiving external feedback on LST deliverables; and living slides have been created for use by LST members to minimise repetitive work and unify messaging when presenting on LISA and LST. Internal links are being supported by streamlining interactions between the various WGs in reporting and presenting their progress and deliverables to both other LST members and the scientific community; this includes a response to questions document that ensures consistent messaging about mission timelines, outcomes, authorship, etc.


\subsection{The Figures of Merit Working Group}


Chaired by AP and AS, with several LST and external members, this WG is concerned with the Figures of Merit (FoM), a set of metrics designed to quantify the mission’s ability to meet its science goals, creating a direct link from instrument specs to science objectives. They provide a key tool for the Performance \& Operations team, particularly during the development phase when technical or financial constraints may require modifications to the instrument design. The FoM enable such changes to be assessed in terms of their impact on the scientific return of the mission. They also allow consideration for points of failure during the mission and the scientific consequences. For this reason, ESA has commissioned the LST to define and implement the FoM. The group is currently reviewing the existing set of FoM, both a static \cite{FoM:2024static} and interactive site \cite{FoM:2024active}, prepared during earlier stages of the mission definition by the LISA Consortium. 
The immediate task is to simplify and streamline these metrics, and to produce a consolidated document with the final version of the FoM (LST version). 
Ultimately, the WG will deliver and maintain an FoM Tool for the Performance \& Operations team, understanding how to best achieve this is also being considered by the WG.


\subsection{The L3 Catalogue Working Group}


This WG, chaired by AL and NK, has been charged with determining the content and format of the Level 3 science catalogue and accompanying LISA data release, which is required to include GW candidates with detection confidence, estimated astrophysical parameters, the overall operational status of the instrument, strain time series, and the residual L1 datastream with candidate sources removed. The final aim is to define a catalogue format which will enable and maximize the scientific output of the mission by the broad community. To do this, the WG is working on identifying functional priorities and design features, e.g. data visualisation to enable preliminary analysis and catalog cross-matching; these will lead into detailed descriptions of tools to be developed for interfacing with the data that will be provided with the data releases. The goal of easy accessibility for all scientists requires considerations of accessibility issues for scientists with no GW specialty. 


\subsection{The Science Topical Panels Working Group}


The goal of the Science Topical Panels (STP) WG, chaired by EMR and ST, is to determine the nature of STPs during the Early Release Science Time of the LISA mission, defined as approximately the first 12 months of data collection after a 3-month period of in-orbit commissioning \cite{LISA:SMP}. The STP WG discusses potential panel topics, team composition (including chairs), required expertise, member responsibilities, interaction with the LISA Collaboration (defined in the SMP as everybody on the author lists) and the scientific community, and how all of these issues feed into the solicitation and timeline procedure for topics and members. The goal of the WG is to generate a reduced set of proposed actions that encapsulate different scenarios for broader LST discussion. The STP WG has produced and presented a draft document to the LST that describes several possible procedures that lead to the formation of STPs. Further streamlining of this document, taking into consideration feedback from the LST, is in progress. In addition, an avenue for receiving feedback from the scientific community is being developed.


\section{Acknowledgements}


The authors would like to thank their fellow members and project scientists of the LISA science team for their invaluable feedback and illuminating discussions, in particular, Ann Hornschemeier Cardiff, Oliver Jennrich and Ira Thorpe. We also thank Maarten van de Meent for the use of Fig.~\ref{fig:parameterSpace}, which he previously produced for~\cite{LISAConsortiumWaveformWorkingGroup:2023arg}. This short review was partially completed while AH and JG were visiting the Institute for Mathematical Sciences, National University of Singapore in 2025. AH is supported by grant PD-034-2023 co-financed by the Govern Balear and the European Social Fund Plus (ESF+) 2021-2027. This work was supported by the Universitat de les Illes Balears (UIB) with funds from the Programa de Foment de la Recerca i la Innovació de la UIB 2024-2026 (supported by the yearly plan of the Tourist Stay Tax ITS2023-086); the Spanish Agencia Estatal de Investigación grants PID2022-138626NB-I00, RED2024-153978-E, RED2024-153735-E, funded by MICIU/AEI/10.13039/501100011033 and the ERDF/EU; and the Comunitat Autònoma de les Illes Balears through the Conselleria d'Educació i Universitats with funds from the European Union - European Regional Development Fund (ERDF) (SINCO2022/18146 - Plataforma HiTech-IAC3-BIO). NK was supported by the Hellenic Foundation for Research and Innovation (H.F.R.I.) under the 4th Call for HFRI Research Projects to support Post-doctoral Researchers (Project Number: 28418). AL acknowledges support from the Centre National des Etudes Spatiales (CNES), CNRS MITI Interdisciplinary research progams and the French Agence Nationale de la Recherche (ANR). AP acknowledges support from the Centre National des Etudes Spatiales (CNES). EMR acknowledge support from the European Research Council (ERC) grant number: 101002511/project acronym: VEGA\_P, as well as NVO VICI grant VI.C.232.099. AS acknowledges the financial support provided under the European Union’s H2020 ERC Advanced Grant ``PINGU'' (Grant Agreement: 101142079). JSK is supported by a UW Global Innovation Fund Research Award. DS is supported by NASA Grant No. 80NSSC24K0437 and NSF Grant No. PHY-2207780. KLS gratefully acknowledges the Mitchell-Heep-Munnerlyn Career Enhancement chair at Texas A\&M University. SRT is supported by an NSF CAREER \#2146016, NSF AST-2307719, NSF NRT-2125764, and NASA LPS-80NSSC26K0342. SRT also acknowledges support from a Chancellor's Faculty Fellowship from Vanderbilt University.  AV acknowledges the support of the UK Space Agency, Grant No. ST/V002813/1 and UKRI971, and of the Royal Society and Wolfson Foundation. CG, AS, and WJW have been supported by the Agenzia Spaziale Italiana (ASI), Accordo n.2024-36-HH.0, ``Attività per la fase B2/C della missione LISA''.  

\section*{References}

\bibliographystyle{iopart-num}
\bibliography{LST}

@article{LISA:2024hlh,
    author = "Colpi, Monica and others",
    collaboration = "LISA",
    title = "{LISA Definition Study Report}",
    primaryClass = "astro-ph.CO",
    month = "2",
    year = "2024",
    note = {\href{https://arxiv.org/abs/2402.07571}{arXiv:2402.07571}}
}

@misc{LISA:SMP,
 author = {{European Space Agency}},
  title = {LISA Science Management Plan},
  year = 2024,
  howpublished = {\url{https://www.cosmos.esa.int/documents/15452792/15452811/LISA-Science-Management-Plan.pdf}},
  note         = {Accessed: 2026-04-21}
}

@article{lpf_noise_2016,
author = "M.~Armano and others",
	title  = "Sub-Femto-g Free Fall for Space-Based Gravitational Wave Observatories:
LISA Pathfinder Results",
	journal= "Phys. Rev. Lett.",
	volume = "116",
	pages  = "231101",
	year   = "2016"	}

@article{lpf_noise_final,
author = "M.~Armano and others",
	title  = "Beyond the Required LISA Free-Fall Performance: New LISA Pathfinder Results
down to 20~$\mu$Hz",
	journal= "Phys. Rev. Lett.",
	volume = "120",
	pages  = "061101",
	year   = "2018"	}

@article{Grimani:2022bkr,
    author = "Grimani, Catia and Villani, Mattia and Fabi, Michele and Cesarini, Andrea and Sabbatini, Federico",
    title = "{Bridging the gap between Monte Carlo simulations and measurements of the LISA Pathfinder test-mass charging for LISA}",
    primaryClass = "astro-ph.HE",
    doi = "10.1051/0004-6361/202243984",
    journal = "Astron. Astrophys.",
    volume = "666",
    pages = "A38",
    year = "2022",
    note = {\href{https://arxiv.org/abs/2208.08849}{arXiv:2208.08849}}
}

@article{Wass:2022igo,
    author = "Wass, P. J. and Sumner, T. J. and Ara{\'u}jo, H. M. and Hollington, D.",
    title = "{Simulating the charging of isolated free-falling masses from TeV to eV energies: Detailed comparison with LISA Pathfinder results}",
    primaryClass = "astro-ph.IM",
    doi = "10.1103/PhysRevD.107.022010",
    journal = "Phys. Rev. D",
    volume = "107",
    number = "2",
    pages = "022010",
    year = "2023",
    note = {\href{https://arxiv.org/abs/2211.09987}{arXiv:2211.09987}}
}

@ARTICLE{2025A&A...700A.102D,
       author = {{Dimiccoli}, F. and {Dolesi}, R. and {Fabi}, M. and {Ferroni}, V. and {Grimani}, C. and {Muratore}, M. and {Sarra}, P. and {Villani}, M. and {Weber}, W.~J.},
        title = "{LISA test-mass charging: Particle flux modeling, Monte Carlo simulations, and induced effects on the sensitivity of the observatory}",
      journal = {Astronomy \& Astrophysics},
         year = 2025,
        month = aug,
       volume = {700},
          eid = {A102},
        pages = {A102},
          doi = {10.1051/0004-6361/202453076},
 primaryClass = {astro-ph.IM},
       adsurl = {https://ui.adsabs.harvard.edu/abs/2025A&A...700A.102D},
    note = {\href{https://arxiv.org/abs/2411.18030}{arXiv:2411.18030}}
}

@article{LISAConsortiumWaveformWorkingGroup:2023arg,
    author = "Afshordi, Niayesh and others",
    collaboration = "LISA Consortium Waveform Working Group",
    title = "{Waveform modelling for the Laser Interferometer Space Antenna}",
    primaryClass = "gr-qc",
    doi = "10.1007/s41114-025-00056-1",
    journal = "Living Rev. Rel.",
    volume = "28",
    number = "1",
    pages = "9",
    year = "2025",
    note = {\href{https://arxiv.org/abs/2311.01300}{arXiv:2311.01300}}
}

@article{LISA:2022yao,
    author = "Seoane, Pau Amaro and others",
    collaboration = "LISA",
    title = "{Astrophysics with the Laser Interferometer Space Antenna}",
    primaryClass = "gr-qc",
    doi = "10.1007/s41114-022-00041-y",
    journal = "Living Rev. Rel.",
    volume = "26",
    number = "1",
    pages = "2",
    year = "2023",
    note = {\href{https://arxiv.org/abs/2203.06016}{arXiv:2203.06016}}
}

@article{LISACosmologyWorkingGroup:2022jok,
    author = "Auclair, Pierre and others",
    collaboration = "LISA Cosmology Working Group",
    title = "{Cosmology with the Laser Interferometer Space Antenna}",
    eprint = "2204.05434",
    archivePrefix = "arXiv",
    primaryClass = "astro-ph.CO",
    reportNumber = "LISA CosWG-22-03, FERMILAB-PUB-22-349-SCD",
    doi = "10.1007/s41114-023-00045-2",
    journal = "Living Rev. Rel.",
    volume = "26",
    number = "1",
    pages = "5",
    year = "2023"
}

@article{LISA:2022kgy,
    author = "Arun, K. G. and others",
    collaboration = "LISA",
    title = "{New horizons for fundamental physics with LISA}",
    eprint = "2205.01597",
    archivePrefix = "arXiv",
    primaryClass = "gr-qc",
    doi = "10.1007/s41114-022-00036-9",
    journal = "Living Rev. Rel.",
    volume = "25",
    number = "1",
    pages = "4",
    year = "2022"
}

@article{Breivik:2017jip,
    author = "Breivik, Katelyn and Kremer, Kyle and Bueno, Michael and Larson, Shane L. and Coughlin, Scott and Kalogera, Vassiliki",
    title = "{Characterizing Accreting Double White Dwarf Binaries with the Laser Interferometer Space Antenna and Gaia}",
    primaryClass = "astro-ph.SR",
    doi = "10.3847/2041-8213/aaaa23",
    journal = "Astrophys. J. Lett.",
    volume = "854",
    number = "1",
    pages = "L1",
    year = "2018",
    note = {\href{https://arxiv.org/abs/1710.08370}{arXiv:1710.08370}}
}

@article{Kupfer:2023nqx,
    author = "Kupfer, Thomas and others",
    title = "{LISA Galactic Binaries with Astrometry from Gaia DR3}",
    primaryClass = "astro-ph.SR",
    doi = "10.3847/1538-4357/ad2068",
    journal = "Astrophys. J.",
    volume = "963",
    number = "2",
    pages = "100",
    year = "2024",
    note = {\href{https://arxiv.org/abs/2302.12719}{arXiv:2302.12719}}
}

@article{Blanchet:2013haa,
    author = "Blanchet, Luc",
    title = "{Post-Newtonian Theory for Gravitational Waves}",
    primaryClass = "gr-qc",
    doi = "10.12942/lrr-2014-2",
    journal = "Living Rev. Rel.",
    volume = "17",
    pages = "2",
    year = "2014",
    note = {\href{https://arxiv.org/abs/1310.1528}{arXiv:1310.1528}}
}

@article{Marsh:2003rd,
    author = "Marsh, Thomas Richard and Nelemans, G. and Steeghs, D.",
    title = "{Mass transfer between double white dwarfs}",
    doi = "10.1111/j.1365-2966.2004.07564.x",
    journal = "Mon. Not. Roy. Astron. Soc.",
    volume = "350",
    pages = "113",
    year = "2004",
    note = {\href{https://arxiv.org/abs/astro-ph/0312577}{arXiv:astro-ph/0312577}}
}

@article{Giacobbo:2018etu,
    author = "Giacobbo, Nicola and Mapelli, Michela",
    title = "{The progenitors of compact-object binaries: impact of metallicity, common envelope and natal kicks}",
    primaryClass = "astro-ph.HE",
    doi = "10.1093/mnras/sty1999",
    journal = "Mon. Not. Roy. Astron. Soc.",
    volume = "480",
    number = "2",
    pages = "2011--2030",
    year = "2018",
    note = {\href{https://arxiv.org/abs/1806.00001}{arXiv:1806.00001}}
}

@article{LIGOScientific:2025rsn,
    author = "Abac, A. G. and others",
    collaboration = "LIGO Scientific, VIRGO, KAGRA",
    title = "{GW231123: A Binary Black Hole Merger with Total Mass 190{\textendash}265 M$_{⊙}$}",
    primaryClass = "astro-ph.HE",
    reportNumber = "DCC: P2500026-v6, DCC: P2500026-v8",
    doi = "10.3847/2041-8213/ae0c9c",
    journal = "Astrophys. J. Lett.",
    volume = "993",
    number = "1",
    pages = "L25",
    year = "2025",
    note = {\href{https://arxiv.org/abs/2507.08219}{arXiv:2507.08219}}
}

@article{Lin:2018dev,
    author = "Lin, Dacheng and others",
    title = "{A luminous X-ray outburst from an intermediate-mass black hole in an off-centre star cluster}",
    eprint = "1806.05692",
    archivePrefix = "arXiv",
    primaryClass = "astro-ph.HE",
    doi = "10.1038/s41550-018-0493-1",
    journal = "Nature Astron.",
    volume = "2",
    number = "8",
    pages = "656--661",
    year = "2018",
    note = {\href{https://arxiv.org/abs/1806.05692}{arXiv:1806.05692}}
}

@ARTICLE{2020ARA&A..58..257G,
       author = {Greene, Jenny E. and Strader, Jay and Ho, Luis C.},
        title = "{Intermediate-Mass Black Holes}",
      journal = {Annu. Rev. Astron. Astrophys.},
         year = 2020,
        month = aug,
       volume = {58},
        pages = {257-312},
          doi = {10.1146/annurev-astro-032620-021835},
 primaryClass = {astro-ph.GA},
       adsurl = {https://ui.adsabs.harvard.edu/abs/2020ARA&A..58..257G},
    note = {\href{https://arxiv.org/abs/1911.09678}{arXiv:1911.09678}}
}

@article{Maiolino:2023bpi,
    author = "Maiolino, Roberto and others",
    title = "{JADES - The diverse population of infant black holes at 4 {\ensuremath{<}} z {\ensuremath{<}} 11: Merging, tiny, poor, but mighty}",
    primaryClass = "astro-ph.GA",
    doi = "10.1051/0004-6361/202347640",
    journal = "Astron. Astrophys.",
    volume = "691",
    pages = "A145",
    year = "2024",
    note = {\href{https://arxiv.org/abs/2308.01230}{arXiv:2308.01230}}
}

@article{Banados:2017unc,
    author = "Banados, Eduardo and others",
    title = "{An 800-million-solar-mass black hole in a significantly neutral Universe at redshift 7.5}",
    eprint = "1712.01860",
    archivePrefix = "arXiv",
    primaryClass = "astro-ph.GA",
    doi = "10.1038/nature25180",
    journal = "Nature",
    volume = "553",
    number = "7689",
    pages = "473--476",
    year = "2018",
    note = {\href{https://arxiv.org/abs/1712.01860}{arXiv:1712.01860}}
}

@article{LIGOScientific:2017ync,
    author = "Abbott, B. P. and others",
    collaboration = "LIGO Scientific, Virgo, Fermi GBM, INTEGRAL, IceCube, AstroSat Cadmium Zinc Telluride Imager Team, IPN, Insight-Hxmt, ANTARES, Swift, AGILE Team, 1M2H Team, Dark Energy Camera GW-EM, DES, DLT40, GRAWITA, Fermi-LAT, ATCA, ASKAP, Las Cumbres Observatory Group, OzGrav, DWF (Deeper Wider Faster Program), AST3, CAASTRO, VINROUGE, MASTER, J-GEM, GROWTH, JAGWAR, CaltechNRAO, TTU-NRAO, NuSTAR, Pan-STARRS, MAXI Team, TZAC Consortium, KU, Nordic Optical Telescope, ePESSTO, GROND, Texas Tech University, SALT Group, TOROS, BOOTES, MWA, CALET, IKI-GW Follow-up, H.E.S.S., LOFAR, LWA, HAWC, Pierre Auger, ALMA, Euro VLBI Team, Pi of Sky, Chandra Team at McGill University, DFN, ATLAS Telescopes, High Time Resolution Universe Survey, RIMAS, RATIR, SKA South Africa/MeerKAT",
    title = "{Multi-messenger Observations of a Binary Neutron Star Merger}",
    primaryClass = "astro-ph.HE",
    reportNumber = "LIGO-P1700294, VIR-0802A-17, FERMILAB-PUB-17-478-A-AE-CD",
    doi = "10.3847/2041-8213/aa91c9",
    journal = "Astrophys. J. Lett.",
    volume = "848",
    number = "2",
    pages = "L12",
    year = "2017",
    note = {\href{https://arxiv.org/abs/1710.05833}{arXiv:1710.05833}}
}

@article{LIGOScientific:2025yae,
    author = "Abac, A. G. and others",
    collaboration = "LIGO Scientific, VIRGO, KAGRA",
    title = "{GWTC-4.0: Methods for Identifying and Characterizing Gravitational-wave Transients}",
    primaryClass = "gr-qc",
    reportNumber = "LIGO-P2400300",
    month = "8",
    year = "2025",
    note = {\href{https://arxiv.org/abs/2508.18081}{arXiv:2508.18081}}
}

@article{Ramos-Buades:2023ehm,
    author = "Ramos-Buades, Antoni and Buonanno, Alessandra and Estell{\'e}s, H{\'e}ctor and Khalil, Mohammed and Mihaylov, Deyan P. and Ossokine, Serguei and Pompili, Lorenzo and Shiferaw, Mahlet",
    title = "{Next generation of accurate and efficient multipolar precessing-spin effective-one-body waveforms for binary black holes}",
    primaryClass = "gr-qc",
    doi = "10.1103/PhysRevD.108.124037",
    journal = "Phys. Rev. D",
    volume = "108",
    number = "12",
    pages = "124037",
    year = "2023",
    note = {\href{https://arxiv.org/abs/2303.18046}{arXiv:2303.18046}}
}

@article{Akcay:2018yyh,
    author = "Akcay, Sarp and Bernuzzi, Sebastiano and Messina, Francesco and Nagar, Alessandro and Ortiz, N{\'e}stor and Rettegno, Piero",
    title = "{Effective-one-body multipolar waveform for tidally interacting binary neutron stars up to merger}",
    primaryClass = "gr-qc",
    doi = "10.1103/PhysRevD.99.044051",
    journal = "Phys. Rev. D",
    volume = "99",
    number = "4",
    pages = "044051",
    year = "2019",
    note = {\href{https://arxiv.org/abs/1812.02744}{arXiv:1812.02744}}
}

@article{Colleoni:2024knd,
    author = "Colleoni, Marta and Vidal, Felip A. Ramis and Garc{\'\i}a-Quir{\'o}s, Cecilio and Ak{\c{c}}ay, Sarp and Bera, Sayantani",
    title = "{Fast frequency-domain gravitational waveforms for precessing binaries with a new twist}",
    primaryClass = "gr-qc",
    doi = "10.1103/PhysRevD.111.104019",
    journal = "Phys. Rev. D",
    volume = "111",
    number = "10",
    pages = "104019",
    year = "2025",
    note = {\href{https://arxiv.org/abs/2412.16721}{arXiv:2412.16721}}
}

@article{Thompson:2023ase,
    author = "Thompson, Jonathan E. and Hamilton, Eleanor and London, Lionel and Ghosh, Shrobana and Kolitsidou, Panagiota and Hoy, Charlie and Hannam, Mark",
    title = "{PhenomXO4a: a phenomenological gravitational-wave model for precessing black-hole binaries with higher multipoles and asymmetries}",
    primaryClass = "gr-qc",
    reportNumber = "LIGO-P2300437",
    doi = "10.1103/PhysRevD.109.063012",
    journal = "Phys. Rev. D",
    volume = "109",
    number = "6",
    pages = "063012",
    year = "2024",
    note = {\href{https://arxiv.org/abs/2312.10025}{arXiv:2312.10025}}
}

@article{Varma:2019csw,
    author = "Varma, Vijay and Field, Scott E. and Scheel, Mark A. and Blackman, Jonathan and Gerosa, Davide and Stein, Leo C. and Kidder, Lawrence E. and Pfeiffer, Harald P.",
    title = "{Surrogate models for precessing binary black hole simulations with unequal masses}",
    primaryClass = "gr-qc",
    doi = "10.1103/PhysRevResearch.1.033015",
    journal = "Phys. Rev. Research.",
    volume = "1",
    pages = "033015",
    year = "2019",
    note = {\href{https://arxiv.org/abs/1905.09300}{arXiv:1905.09300}}
}

@article{Hopman:2005vr,
    author = "Hopman, Clovis and Alexander, Tal",
    title = "{The Orbital statistics of stellar inspiral and relaxation near a massive black hole: Characterizing gravitational wave sources}",
    doi = "10.1086/431475",
    journal = "Astrophys. J.",
    volume = "629",
    pages = "362--372",
    year = "2005",
    note = {\href{https://arxiv.org/abs/astro-ph/0503672}{arXiv:astro-ph/0503672}}
}

@article{ColemanMiller:2005rm,
    author = "Coleman Miller, M. and Freitag, Marc and Hamilton, Douglas P. and Lauburg, Vanessa M.",
    title = "{Binary encounters with supermassive black holes: Zero-eccentricity LISA events}",
    doi = "10.1086/497335",
    journal = "Astrophys. J. Lett.",
    volume = "631",
    pages = "L117--L120",
    year = "2005",
    note = {\href{https://arxiv.org/abs/astro-ph/0507133}{arXiv:astro-ph/0507133}}
}

@article{Volonteri:2004cf,
    author = "Volonteri, Marta and Madau, Piero and Quataert, Eliot and Rees, Martin J.",
    title = "{The Distribution and cosmic evolution of massive black hole spins}",
    doi = "10.1086/426858",
    journal = "Astrophys. J.",
    volume = "620",
    pages = "69--77",
    year = "2005",
    note = {\href{https://arxiv.org/abs/astro-ph/0410342}{arXiv:astro-ph/0410342}}
}

@article{Duque:2025yfm,
    author = "Duque, Francisco and Sberna, Laura and Spiers, Andrew and Vicente, Rodrigo",
    title = "{Extreme-mass-ratio inspirals in relativistic accretion discs}",
    primaryClass = "gr-qc",
    doi = "10.1103/jcv5-ssfd",
    journal = "Phys. Rev. D",
    volume = "113",
    number = "8",
    pages = "084028",
    year = "2026",
    note = {\href{https://arxiv.org/abs/2510.02433}{arXiv:2510.02433}}
}

@article{HegadeKR:2025rpr,
    author = "Hegade K. R., Abhishek and Gammie, Charles F. and Yunes, Nicol{\'a}s",
    title = "{Relativistic treatment of accretion disk torques on extreme mass ratio inspirals around spinning black holes}",
    primaryClass = "gr-qc",
    doi = "10.1103/g83s-jdld",
    journal = "Phys. Rev. D",
    volume = "112",
    number = "12",
    pages = "124068",
    year = "2025",
    note = {\href{https://arxiv.org/abs/2510.03564}{arXiv:2510.03564}}
}

@article{Pan:2023wau,
    author = "Pan, Zhen and Yang, Huan and Bernard, Laura and Bonga, B{\'e}atrice",
    title = "{Resonant dynamics of extreme mass-ratio inspirals in a perturbed Kerr spacetime}",
    primaryClass = "gr-qc",
    doi = "10.1103/PhysRevD.108.104026",
    journal = "Phys. Rev. D",
    volume = "108",
    number = "10",
    pages = "104026",
    year = "2023",
    note = {\href{https://arxiv.org/abs/2306.06576}{arXiv:2306.06576}}
}

@article{Wardell:2021fyy,
    author = "Wardell, Barry and Pound, Adam and Warburton, Niels and Miller, Jeremy and Durkan, Leanne and Le Tiec, Alexandre",
    title = "{Gravitational Waveforms for Compact Binaries from Second-Order Self-Force Theory}",
    primaryClass = "gr-qc",
    doi = "10.1103/PhysRevLett.130.241402",
    journal = "Phys. Rev. Lett.",
    volume = "130",
    number = "24",
    pages = "241402",
    year = "2023",
    note = {\href{https://arxiv.org/abs/2112.12265}{arXiv:2112.12265}}
}

@article{Mathews:2021rod,
    author = "Mathews, Josh and Pound, Adam and Wardell, Barry",
    title = "{Self-force calculations with a spinning secondary}",
    primaryClass = "gr-qc",
    doi = "10.1103/PhysRevD.105.084031",
    journal = "Phys. Rev. D",
    volume = "105",
    number = "8",
    pages = "084031",
    year = "2022",
    note = {\href{https://arxiv.org/abs/2112.13069}{arXiv:2112.13069}}
}

@article{Mathews:2025txc,
    author = "Mathews, Josh and Wardell, Barry and Pound, Adam and Warburton, Niels",
    title = "{Postadiabatic self-force waveforms: Slowly spinning primary and precessing secondary}",
    primaryClass = "gr-qc",
    doi = "10.1103/ph3p-mscl",
    journal = "Phys. Rev. D",
    volume = "113",
    number = "6",
    pages = "064034",
    year = "2026",
    note = {\href{https://arxiv.org/abs/2510.16113}{arXiv:2510.16113}}
}

@article{Kuchler:2025hwx,
    author = {K{\"u}chler, Lorenzo and Comp{\`e}re, Geoffrey and Pound, Adam},
    title = "{Self-force framework for merger-ringdown waveforms}",
    primaryClass = "gr-qc",
    doi = "10.1088/1361-6382/ae2b44",
    journal = "Class. Quant. Grav.",
    volume = "43",
    number = "1",
    pages = "015018",
    year = "2026",
    note = {\href{https://arxiv.org/abs/2506.02189}{arXiv:2506.02189}}
}

@article{Honet:2025dho,
    author = {Honet, Lo{\"\i}c and K{\"u}chler, Lorenzo and Pound, Adam and Comp{\`e}re, Geoffrey},
    title = "{Transition-to-plunge self-force waveforms with a spinning primary}",
    primaryClass = "gr-qc",
    doi = "10.1103/sq6y-qv8h",
    journal = "Phys. Rev. D",
    volume = "113",
    number = "4",
    pages = "044051",
    year = "2026",
    note = {\href{https://arxiv.org/abs/2510.13958}{arXiv:2510.13958}}
}

@article{Burke:2023lno,
    author = "Burke, Ollie and Piovano, Gabriel Andres and Warburton, Niels and Lynch, Philip and Speri, Lorenzo and Kavanagh, Chris and Wardell, Barry and Pound, Adam and Durkan, Leanne and Miller, Jeremy",
    title = "{Assessing the importance of first postadiabatic terms for small-mass-ratio binaries}",
    primaryClass = "gr-qc",
    doi = "10.1103/PhysRevD.109.124048",
    journal = "Phys. Rev. D",
    volume = "109",
    number = "12",
    pages = "124048",
    year = "2024",
    note = {\href{https://arxiv.org/abs/2310.08927}{arXiv:2310.08927}}
}

@article{Chapman-Bird:2025xtd,
    author = "Chapman-Bird, Christian E. A. and others",
    title = "{Efficient waveforms for asymmetric-mass eccentric equatorial inspirals into rapidly spinning black holes}",
    primaryClass = "gr-qc",
    doi = "10.1103/scbp-75pf",
    journal = "Phys. Rev. D",
    volume = "112",
    number = "10",
    pages = "104023",
    year = "2025",
    note = {\href{https://arxiv.org/abs/2506.09470}{arXiv:2506.09470}}
}

@article{Wittek:2024pis,
    author = "Wittek, Nikolas A. and Barack, Leor and Pfeiffer, Harald P. and Pound, Adam and Deppe, Nils and Kidder, Lawrence E. and Macedo, Alexandra and Nelli, Kyle C. and Throwe, William and Vu, Nils L.",
    title = "{Relieving Scale Disparity in Binary Black Hole Simulations}",
    primaryClass = "gr-qc",
    doi = "10.1103/kskl-8dcj",
    journal = "Phys. Rev. Lett.",
    volume = "134",
    number = "25",
    pages = "251402",
    year = "2025",
    note = {\href{https://arxiv.org/abs/2410.22290}{arXiv:2410.22290}}
}

@article{Honet:2025lmk,
    author = {Honet, Lo{\"\i}c and Mathews, Josh and Comp{\`e}re, Geoffrey and Pound, Adam and Wardell, Barry and Piovano, Gabriel Andres and van de Meent, Maarten and Warburton, Niels},
    title = "{Spin-aligned inspiral waveforms from self-force and post-Newtonian theory}",
    primaryClass = "gr-qc",
    month = "10",
    year = "2025",
    note = {\href{https://arxiv.org/abs/2510.16112}{arXiv:2510.16112}}
}

@article{LIGOScientific:2025pvj,
    author = "Abac, A. G. and others",
    collaboration = "LIGO Scientific, VIRGO, KAGRA",
    title = "{GWTC-4.0: Population Properties of Merging Compact Binaries}",
    primaryClass = "astro-ph.HE",
    reportNumber = "LIGO-P2400004",
    month = "8",
    year = "2025",
    note = {\href{https://arxiv.org/abs/2508.18083}{arXiv:2508.18083}}
}

@article{LIGOScientific:2020iuh,
    author = "Abbott, R. and others",
    collaboration = "LIGO Scientific, Virgo",
    title = "{GW190521: A Binary Black Hole Merger with a Total Mass of $150  M_{\odot}$}",
    primaryClass = "gr-qc",
    doi = "10.1103/PhysRevLett.125.101102",
    journal = "Phys. Rev. Lett.",
    volume = "125",
    number = "10",
    pages = "101102",
    year = "2020",
    note = {\href{https://arxiv.org/abs/2009.01075}{arXiv:2009.01075}}
}

@article{Woosley:2021xba,
    author = "Woosley, S. E. and Heger, Alexander",
    title = "{The Pair-Instability Mass Gap for Black Holes}",
    primaryClass = "astro-ph.SR",
    doi = "10.3847/2041-8213/abf2c4",
    journal = "Astrophys. J. Lett.",
    volume = "912",
    number = "2",
    pages = "L31",
    year = "2021",
    note = {\href{https://arxiv.org/abs/2103.07933}{arXiv:2103.07933}}
}

@article{Gerosa:2021mno,
    author = "Gerosa, Davide and Fishbach, Maya",
    title = "{Hierarchical mergers of stellar-mass black holes and their gravitational-wave signatures}",
    primaryClass = "astro-ph.HE",
    doi = "10.1038/s41550-021-01398-w",
    journal = "Nature Astron.",
    volume = "5",
    number = "8",
    pages = "749--760",
    year = "2021",
    note = {\href{https://arxiv.org/abs/2105.03439}{arXiv:2105.03439}}
}

@article{Tagawa:2020qll,
    author = "Tagawa, Hiromichi and Kocsis, Bence and Haiman, Zoltan and Bartos, Imre and Omukai, Kazuyuki and Samsing, Johan",
    title = "{Mass-gap Mergers in Active Galactic Nuclei}",
    primaryClass = "astro-ph.HE",
    doi = "10.3847/1538-4357/abd555",
    journal = "Astrophys. J.",
    volume = "908",
    number = "2",
    pages = "194",
    year = "2021",
    note = {\href{https://arxiv.org/abs/2012.00011}{arXiv:2012.00011}}
}

@article{Bartos:2025pkv,
    author = "Bartos, Imre and Haiman, Zoltan",
    title = "{Accretion is All You Need: Black Hole Spin Alignment in Merger GW231123 Indicates Accretion Pathway}",
    primaryClass = "astro-ph.HE",
    doi = "10.3847/2041-8213/ae2bff",
    journal = "Astrophys. J. Lett.",
    volume = "996",
    number = "2",
    pages = "L44",
    year = "2026",
    note = {\href{https://arxiv.org/abs/2508.08558}{arXiv:2508.08558}}
}

@article{Inayoshi:2017hgw,
    author = "Inayoshi, Kohei and Tamanini, Nicola and Caprini, Chiara and Haiman, Zolt{\'a}n",
    title = "{Probing stellar binary black hole formation in galactic nuclei via the imprint of their center of mass acceleration on their gravitational wave signal}",
    primaryClass = "astro-ph.HE",
    doi = "10.1103/PhysRevD.96.063014",
    journal = "Phys. Rev. D",
    volume = "96",
    number = "6",
    pages = "063014",
    year = "2017",
    note = {\href{https://arxiv.org/abs/1702.06529}{arXiv:1702.06529}}
}

@article{Baker:2022eiz,
    author = "Baker, Tessa and Barausse, Enrico and Chen, Anson and de Rham, Claudia and Pieroni, Mauro and Tasinato, Gianmassimo",
    title = "{Testing gravitational wave propagation with multiband detections}",
    primaryClass = "gr-qc",
    doi = "10.1088/1475-7516/2023/03/044",
    journal = "JCAP",
    volume = "03",
    pages = "044",
    year = "2023",
    note = {\href{https://arxiv.org/abs/2209.14398}{arXiv:2209.14398}}
}

@article{Berti:2005ys,
    author = "Berti, Emanuele and Cardoso, Vitor and Will, Clifford M.",
    title = "{On gravitational-wave spectroscopy of massive black holes with the space interferometer LISA}",
    doi = "10.1103/PhysRevD.73.064030",
    journal = "Phys. Rev. D",
    volume = "73",
    pages = "064030",
    year = "2006",
    note = {\href{https://arxiv.org/abs/gr-qc/0512160}{arXiv:gr-qc/0512160}}
}

@article{Toubiana:2023cwr,
    author = "Toubiana, Alexandre and Pompili, Lorenzo and Buonanno, Alessandra and Gair, Jonathan R. and Katz, Michael L.",
    title = "{Measuring source properties and quasinormal mode frequencies of heavy massive black-hole binaries with LISA}",
    primaryClass = "gr-qc",
    doi = "10.1103/PhysRevD.109.104019",
    journal = "Phys. Rev. D",
    volume = "109",
    number = "10",
    pages = "104019",
    year = "2024",
    note = {\href{https://arxiv.org/abs/2307.15086}{arXiv:2307.15086}}
}

@article{Babak:2017tow,
    author = "Babak, Stanislav and Gair, Jonathan and Sesana, Alberto and Barausse, Enrico and Sopuerta, Carlos F. and Berry, Christopher P. L. and Berti, Emanuele and Amaro-Seoane, Pau and Petiteau, Antoine and Klein, Antoine",
    title = "{Science with the space-based interferometer LISA. V: Extreme mass-ratio inspirals}",
    primaryClass = "gr-qc",
    doi = "10.1103/PhysRevD.95.103012",
    journal = "Phys. Rev. D",
    volume = "95",
    number = "10",
    pages = "103012",
    year = "2017",
    note = {\href{https://arxiv.org/abs/1703.09722}{arXiv:1703.09722}}
}

@article{Tinto:2010hz,
    author = "Tinto, Massimo and da Silva Alves, Marcio Eduardo",
    title = "{LISA Sensitivities to Gravitational Waves from Relativistic Metric Theories of Gravity}",
    primaryClass = "gr-qc",
    doi = "10.1103/PhysRevD.82.122003",
    journal = "Phys. Rev. D",
    volume = "82",
    pages = "122003",
    year = "2010",
    note = {\href{https://arxiv.org/abs/1010.1302}{arXiv:1010.1302}}
}

@article{Barausse:2016eii,
    author = "Barausse, Enrico and Yunes, Nicol{\'a}s and Chamberlain, Katie",
    title = "{Theory-Agnostic Constraints on Black-Hole Dipole Radiation with Multiband Gravitational-Wave Astrophysics}",
    primaryClass = "gr-qc",
    doi = "10.1103/PhysRevLett.116.241104",
    journal = "Phys. Rev. Lett.",
    volume = "116",
    number = "24",
    pages = "241104",
    year = "2016",
    note = {\href{https://arxiv.org/abs/1603.04075}{arXiv:1603.04075}}
}

@article{Carson:2019rda,
    author = "Carson, Zack and Yagi, Kent",
    title = "{Multi-band gravitational wave tests of general relativity}",
    primaryClass = "gr-qc",
    doi = "10.1088/1361-6382/ab5c9a",
    journal = "Class. Quant. Grav.",
    volume = "37",
    number = "2",
    pages = "02LT01",
    year = "2020",
    note = {\href{https://arxiv.org/abs/1905.13155}{arXiv:1905.13155}}
}

@article{deRham:2018red,
    author = "de Rham, Claudia and Melville, Scott",
    title = "{Gravitational Rainbows: LIGO and Dark Energy at its Cutoff}",
    primaryClass = "hep-th",
    reportNumber = "Imperial/TP/2018/CdR/02",
    doi = "10.1103/PhysRevLett.121.221101",
    journal = "Phys. Rev. Lett.",
    volume = "121",
    number = "22",
    pages = "221101",
    year = "2018",
    note = {\href{https://arxiv.org/abs/11806.09417}{arXiv:1806.09417}}
}

@article{Perkins:2020tra,
    author = "Perkins, Scott E. and Yunes, Nicol{\'a}s and Berti, Emanuele",
    title = "{Probing Fundamental Physics with Gravitational Waves: The Next Generation}",
    primaryClass = "gr-qc",
    doi = "10.1103/PhysRevD.103.044024",
    journal = "Phys. Rev. D",
    volume = "103",
    number = "4",
    pages = "044024",
    year = "2021",
    note = {\href{https://arxiv.org/abs/2010.09010}{arXiv:2010.09010}}
}

@article{LISACosmologyWorkingGroup:2019mwx,
    author = "Belgacem, Enis and others",
    collaboration = "LISA Cosmology Working Group",
    title = "{Testing modified gravity at cosmological distances with LISA standard sirens}",
    primaryClass = "astro-ph.CO",
    reportNumber = "LISA CosWG-19-01; IFT-UAM-CSIC-19-79, LISA CosWG-19-01",
    doi = "10.1088/1475-7516/2019/07/024",
    journal = "JCAP",
    volume = "07",
    pages = "024",
    year = "2019",
    note = {\href{https://arxiv.org/abs/1906.01593}{arXiv:1906.01593}}
}

@article{Thrane:2013oya,
    author = "Thrane, Eric and Romano, Joseph D.",
    title = "{Sensitivity curves for searches for gravitational-wave backgrounds}",
    primaryClass = "astro-ph.IM",
    doi = "10.1103/PhysRevD.88.124032",
    journal = "Phys. Rev. D",
    volume = "88",
    number = "12",
    pages = "124032",
    year = "2013",
    note = {\href{https://arxiv.org/abs/1310.5300}{arXiv:1310.5300}}
}

@article{EPTA:2023xxk,
    author = "Antoniadis, J. and others",
    collaboration = "EPTA, InPTA",
    title = "{The second data release from the European Pulsar Timing Array - IV. Implications for massive black holes, dark matter, and the early Universe}",
    primaryClass = "astro-ph.CO",
    doi = "10.1051/0004-6361/202347433",
    journal = "Astron. Astrophys.",
    volume = "685",
    pages = "A94",
    year = "2024",
    note = {\href{https://arxiv.org/abs/2306.16227}{arXiv:2306.16227}}
}

@article{Schutz:1986gp,
    author = "Schutz, Bernard F.",
    title = "{Determining the Hubble Constant from Gravitational Wave Observations}",
    doi = "10.1038/323310a0",
    journal = "Nature",
    volume = "323",
    pages = "310--311",
    year = "1986"
}

@article{Mangiagli:2022niy,
    author = "Mangiagli, Alberto and Caprini, Chiara and Volonteri, Marta and Marsat, Sylvain and Vergani, Susanna and Tamanini, Nicola and Inchausp{\'e}, Henri",
    title = "{Massive black hole binaries in LISA: Multimessenger prospects and electromagnetic counterparts}",
    primaryClass = "astro-ph.HE",
    doi = "10.1103/PhysRevD.106.103017",
    journal = "Phys. Rev. D",
    volume = "106",
    number = "10",
    pages = "103017",
    year = "2022",
    note = {\href{https://arxiv.org/abs/2207.10678}{arXiv:2207.10678}}
}

@article{Mangiagli:2023ize,
    author = "Mangiagli, Alberto and Caprini, Chiara and Marsat, Sylvain and Speri, Lorenzo and Caldwell, Robert R. and Tamanini, Nicola",
    title = "{Massive black hole binaries in LISA: Constraining cosmological parameters at high redshifts}",
    primaryClass = "astro-ph.CO",
    doi = "10.1103/PhysRevD.111.083043",
    journal = "Phys. Rev. D",
    volume = "111",
    number = "8",
    pages = "083043",
    year = "2025",
    note = {\href{https://arxiv.org/abs/2312.04632}{arXiv:2312.04632}}
}

@article{Speri:2020hwc,
    author = "Speri, Lorenzo and Tamanini, Nicola and Caldwell, Robert R. and Gair, Jonathan R. and Wang, Benjamin",
    title = "{Testing the Quasar Hubble Diagram with LISA Standard Sirens}",
    primaryClass = "astro-ph.CO",
    doi = "10.1103/PhysRevD.103.083526",
    journal = "Phys. Rev. D",
    volume = "103",
    number = "8",
    pages = "083526",
    year = "2021",
    note = {\href{https://arxiv.org/abs/2010.09049}{arXiv:2010.09049}}
}

@article{Laghi:2021pqk,
    author = "Laghi, Danny and Tamanini, Nicola and Del Pozzo, Walter and Sesana, Alberto and Gair, Jonathan and Babak, Stanislav and Izquierdo-Villalba, David",
    title = "{Gravitational-wave cosmology with extreme mass-ratio inspirals}",
    primaryClass = "astro-ph.CO",
    doi = "10.1093/mnras/stab2741",
    journal = "Mon. Not. Roy. Astron. Soc.",
    volume = "508",
    number = "3",
    pages = "4512--4531",
    year = "2021",
    note = {\href{https://arxiv.org/abs/2102.01708}{arXiv:2102.01708}}
}

@article{Boileau:2021sni,
    author = "Boileau, Guillaume and Lamberts, Astrid and Christensen, Nelson and Cornish, Neil J. and Meyer, Renate",
    title = "{Spectral separation of the stochastic gravitational-wave background for LISA in the context of a modulated Galactic foreground}",
    eprint = "2105.04283",
    archivePrefix = "arXiv",
    primaryClass = "gr-qc",
    doi = "10.1093/mnras/stab2575",
    journal = "Mon. Not. Roy. Astron. Soc.",
    volume = "508",
    number = "1",
    pages = "803--826",
    year = "2021",
    note = {\href{https://arxiv.org/abs/2105.04283}{arXiv:2105.04283}}
}

@article{Boileau:2025jkv,
    author = "Boileau, Guillaume and Bruel, Tristan and Toubiana, Alexandre and Lamberts, Astrid and Christensen, Nelson",
    title = "{Gravitational-wave background from extragalactic double white dwarfs for LISA}",
    primaryClass = "gr-qc",
    doi = "10.1051/0004-6361/202556052",
    journal = "Astron. Astrophys.",
    volume = "702",
    pages = "A246",
    year = "2025",
    note = {\href{https://arxiv.org/abs/2506.18390}{arXiv:2506.18390}}
}

@article{Karnesis:2021tsh,
    author = "Karnesis, Nikolaos and Babak, Stanislav and Pieroni, Mauro and Cornish, Neil and Littenberg, Tyson",
    title = "{Characterization of the stochastic signal originating from compact binary populations as measured by LISA}",
    primaryClass = "astro-ph.IM",
    doi = "10.1103/PhysRevD.104.043019",
    journal = "Phys. Rev. D",
    volume = "104",
    number = "4",
    pages = "043019",
    year = "2021",
    note = {\href{https://arxiv.org/abs/2103.14598}{arXiv:2103.14598}}
}

@article{Babak:2023lro,
    author = "Babak, Stanislav and Caprini, Chiara and Figueroa, Daniel G. and Karnesis, Nikolaos and Marcoccia, Paolo and Nardini, Germano and Pieroni, Mauro and Ricciardone, Angelo and Sesana, Alberto and Torrado, Jes{\'u}s",
    title = "{Stochastic gravitational wave background from stellar origin binary black holes in LISA}",
    primaryClass = "astro-ph.CO",
    doi = "10.1088/1475-7516/2023/08/034",
    journal = "JCAP",
    volume = "08",
    pages = "034",
    year = "2023",
    note = {\href{https://arxiv.org/abs/2304.06368}{arXiv:2304.06368}}
}

@article{KAGRA:2021duu,
    author = "Abbott, R. and others",
    collaboration = "KAGRA, VIRGO, LIGO Scientific",
    title = "{Population of Merging Compact Binaries Inferred Using Gravitational Waves through GWTC-3}",
    primaryClass = "astro-ph.HE",
    reportNumber = "LIGO-P2100239 ; Data release: https://zenodo.org/record/5655785, LIGO-P2100239",
    doi = "10.1103/PhysRevX.13.011048",
    journal = "Phys. Rev. X",
    volume = "13",
    number = "1",
    pages = "011048",
    year = "2023",
    note = {\href{https://arxiv.org/abs/2111.03634}{arXiv:2111.03634}}
}

@article{Bonetti:2020jku,
    author = "Bonetti, Matteo and Sesana, Alberto",
    title = "{Gravitational wave background from extreme mass ratio inspirals}",
    primaryClass = "astro-ph.GA",
    doi = "10.1103/PhysRevD.102.103023",
    journal = "Phys. Rev. D",
    volume = "102",
    number = "10",
    pages = "103023",
    year = "2020",
    note = {\href{https://arxiv.org/abs/2007.14403}{arXiv:2007.14403}}
}

@article{Caprini:2019egz,
    author = "Caprini, Chiara and others",
    title = "{Detecting gravitational waves from cosmological phase transitions with LISA: an update}",
    primaryClass = "astro-ph.CO",
    reportNumber = "DESY-19-159, IPPP/19/27, HIP-2019-14/TH, MITP/19-066, IFT-UAM/CSIC-19-139",
    doi = "10.1088/1475-7516/2020/03/024",
    journal = "JCAP",
    volume = "03",
    pages = "024",
    year = "2020",
    note = {\href{https://arxiv.org/abs/1910.13125}{arXiv:1910.13125}}
}

@article{Auclair:2019wcv,
    author = "Auclair, Pierre and others",
    title = "{Probing the gravitational wave background from cosmic strings with LISA}",
    primaryClass = "astro-ph.CO",
    doi = "10.1088/1475-7516/2020/04/034",
    journal = "JCAP",
    volume = "04",
    pages = "034",
    year = "2020",
    note = {\href{https://arxiv.org/abs/1909.00819}{arXiv:1909.00819}}
}

@article{Bartolo:2018rku,
    author = "Bartolo, N. and De Luca, V. and Franciolini, G. and Peloso, M. and Racco, D. and Riotto, A.",
    title = "{Testing primordial black holes as dark matter with LISA}",
    eprint = "1810.12224",
    archivePrefix = "arXiv",
    primaryClass = "astro-ph.CO",
    doi = "10.1103/PhysRevD.99.103521",
    journal = "Phys. Rev. D",
    volume = "99",
    number = "10",
    pages = "103521",
    year = "2019",
    note = {\href{https://arxiv.org/abs/1810.12224}{arXiv:1810.12224}}
}

@article{Damour:2000wa,
    author = "Damour, Thibault and Vilenkin, Alexander",
    title = "{Gravitational wave bursts from cosmic strings}",
    reportNumber = "IHES-P-00-32",
    doi = "10.1103/PhysRevLett.85.3761",
    journal = "Phys. Rev. Lett.",
    volume = "85",
    pages = "3761--3764",
    year = "2000",
    note = {\href{https://arxiv.org/abs/gr-qc/0004075}{arXiv:gr-qc/0004075}}
}

@article{Damour:2001bk,
    author = "Damour, Thibault and Vilenkin, Alexander",
    title = "{Gravitational wave bursts from cusps and kinks on cosmic strings}",
    eprint = "gr-qc/0104026",
    archivePrefix = "arXiv",
    reportNumber = "IHES-P-01-15",
    doi = "10.1103/PhysRevD.64.064008",
    journal = "Phys. Rev. D",
    volume = "64",
    pages = "064008",
    year = "2001",
    note = {\href{https://arxiv.org/abs/gr-qc/0104026}{arXiv:gr-qc/0104026}}
}

@article{Caprini:2025mfr,
    author = "Caprini, Chiara and Heffernan, Anna and Brito, Richard and Franciolini, Gabriele and Nardini, Germano and Tamanini, Nicola and Steer, Dani{\`e}le",
    collaboration = "LISA Science",
    title = "{Science of the LISA mission: A Summary for the European Strategy for Particle Physics}",
    primaryClass = "gr-qc",
    month = "7",
    year = "2025",
    note = {\href{https://arxiv.org/abs/2507.05130}{arXiv:2507.05130}}
}

@misc{FoM:2024static,
  author = "{LISA Consortium}",
  title = "Figures of Merit (Static)",
  year = "2024",
  note = "\href{https://wiki-lisa.in2p3.fr/fom-sites/dc\_82\_fmin\_1e-4/site/}{https://wiki-lisa.in2p3.fr/fom-sites/dc\_82\_fmin\_1e-4/site/}"
}

@misc{FoM:2024active,
  author = "{LISA Consortium}",
  title = "Figures of Merit (Interactive)",
  year = "2024",
  note = "\href{https://lisa-science-explorer.in2p3.fr/}{https://lisa-science-explorer.in2p3.fr/}"
}

\end{document}